# Attention-Guided Fair AI Modeling for Skin Cancer Diagnosis

Mingcheng Zhu[1,2,3#], Mingxuan Liu[2,3#], Han Yuan[2,3], Yilin Ning[2,3], Zhiyao Luo[1], Tingting Zhu[1], Nan Liu[2,3,4,5,6*]

[1] Department of Engineering Science, University of Oxford, Oxford, United Kingdom
[2] Center for Quantitative Medicine, Duke-NUS Medical School, Singapore, Singapore
[3] Duke-NUS AI + Medical Sciences Initiative, Duke-NUS Medical School, Singapore, Singapore
[4] Department of Biostatistics and Bioinformatics, Duke University, Durham, NC, USA
[5] Pre-hospital & Emergency Research Centre, Health Services Research & Population Health, Duke-NUS Medical School, Singapore, Singapore
[6] NUS Artificial Intelligence Institute, National University of Singapore, Singapore, Singapore

[#] These authors contributed equally

* Corresponding author: Nan Liu, Centre for Quantitative Medicine, Duke-NUS Medical School, 8 College Road, Singapore 169857, Singapore. Phone: +65 6601 6503. Email: liu.nan@duke-nus.edu.sg

## Abstract

Artificial intelligence (AI) has shown remarkable promise in dermatology, offering accurate and non-invasive diagnosis of skin cancer. While extensive research has addressed skin tone-related bias, gender bias in dermatologic AI remains underexplored, leading to unequal care and reinforcing existing gender disparities. In this study, we developed LesionAttn, a fairness-aware algorithm that integrates clinical knowledge into model design by directing attention toward lesion regions, mirroring the diagnostic focus of clinicians. Combined with Pareto-frontier optimization for dual-objective model selection, LesionAttn balances fairness and predictive accuracy. Validated on two large-scale dermatological datasets, LesionAttn significantly mitigates gender bias while maintaining high diagnostic performance, outperforming existing bias mitigation algorithms. Our study highlights the potential of embedding clinical knowledge into AI development to advance both model performance and fairness, and further to foster interdisciplinary collaboration between clinicians and AI developers.

# Introduction

Skin cancer poses a major global health concern, with more than 6.7 million malignant cases reported annually worldwide. Its incidence has risen sharply over the past two decades,[1,2] highlighting the need for efficient and accurate diagnostic tools. Conventional diagnosis often relies on invasive biopsies that are time-consuming and uncomfortable for patients.[3] In response, Artificial Intelligence (AI) has emerged as a transformative solution, enabling non-invasive skin cancer detection directly from dermoscopic images.[4] These models now achieve diagnostic accuracy comparable to or exceeding that of expert dermatologists[5] and are increasingly integrated into clinical practice.[6,7]

Despite their promise, dermatologic AI models remain vulnerable to algorithmic bias, raising concerns about unequal care across patient subgroups. Prior research has primarily focused on disparities across racial and skin tone groups, showing that models trained predominantly on lighter skin often underperform on darker tones.[8-12] These skin tone-related biases risk exacerbating existing inequities in dermatologic outcomes.[13] In contrast, gender bias in AI-based skin cancer diagnosis remains relatively underexplored, perhaps because gender-specific differences—unlike visible variations in skin tone—are more subtle and less readily recognized.[14,15]

Gender disparities in skin cancer detection may stem from both biological and behavioral differences. These include variations in skin texture, hormonal and vascular characteristics[16,17], as well as gendered patterns in sun protection and skin care practices[18,19]. Such factors could lead to uneven performance of AI models across genders, perpetuating diagnostic inequities. Yet, few studies have systematically investigated and mitigated gender bias in dermatologic AI.

The effectiveness of existing bias mitigation algorithms in skin cancer diagnosis remains uncertain. Most were developed and validated on non-medical datasets and may not generalize to dermatologic image analysis[9,13,20], where domain-specific challenges such as variable lighting, occlusion, and ill-defined lesion boundaries complicate bias mitigation.[21] They also usually lack interpretability, modifying model parameters in ways that are difficult to trace or explain, which limits clinical acceptance and trust.[22] Moreover, many achieve fairness at the expense of diagnostic accuracy—for example, by enforcing similarity in latent feature distributions across demographic subgroups, thereby discarding clinically meaningful information.[23,24] Such trade-offs pose a major barrier to the clinical translation of bias mitigation techniques in dermatologic AI.

To address this gap, we developed LesionAttn, a fairness-aware algorithm that guides model attention toward clinically relevant lesion regions and employs Pareto-frontier optimization to balance diagnostic performance and fairness. By embedding clinical knowledge into model design, LesionAttn enhances lesion-focused reasoning, reduces reliance on spurious background features, and mitigates gender bias without sacrificing performance. These findings offer an interpretable and clinically grounded approach to mitigating gender bias in dermatologic AI and underscore the potential of domain-informed AI to advance both fairness and reliability in clinical settings.

# Results

In this study, we proposed LesionAttn, a fairness-aware algorithm that guides attention toward clinically relevant lesion areas while optimizing both performance and fairness via Pareto-frontier model selection (Figure 1). See more details about LesionAttn in the Methods section under "LesionAttn Algorithm for bias mitigation". LesionAttn was evaluated against existing fairness algorithms across two large-scale public dermatologic datasets—the Human Against Machine dataset (HAM)[25] and the BCN20000 dataset (BCN)[26] for skin malignancy prediction. We further compared LesionAttn with a static lesion-masking approach, highlighting the importance of maintaining spatial context for both diagnostic performance and fairness.

Throughout this study, unfairness (bias) was quantified using Equalized Odds (EO), defined as the larger disparity between groups in either true positive rates (TPR) or false positive rates (FPR). We also denoted the difference in TPR between male and female groups as EO[TP], and the difference in FPR as EO[FP]. Model performance was evaluated using the areas under the Receiver Operating Characteristic (AUROC) and precision-recall curves (AUPRC) for predictive performance and EO for fairness.

### Gender bias mitigation using LesionAttn

We discovered that the Residual Attention Neural Network (RANN) model, as a widely adopted model architecture in skin cancer diagnosis[27-29], displayed marked gender disparity in both internal (HAM testing set) and external (BCN dataset) cohorts. The RANN model achieved high discrimination performance on HAM (AUROC: 0.903 [0.897-0.909]; AUPRC: 0.698 [0.668-0.727]) and medium performance on the BCN (AUROC: 0.723 [0.696-0.750]; AUPRC: 0.763 [0.742-0.784]) datasets. The corresponding ROC and PRC curves are provided in Supplementary eFigure 1. As shown in Figure 2A, in both datasets, the male subgroup exhibited a significantly higher TPR and lower TNR (i.e., 1-FPR), with statistical significance compared to the female subgroups.

LesionAttn consistently outperformed the baseline model (i.e., RANN) in reducing gender bias while maintaining high predictive performance. In the HAM dataset, LesionAttn

reduced EO by 40.4%, from 0.109 (0.068-0.150) to 0.065 (0.041-0.089), with an AUROC of 0.909 (0.904-0.913) and AUPRC of 0.719 (0.696-0.742), higher than the baseline model. In the BCN dataset, LesionAttn reduced EO by 23.5%, from 0.162 (0.146–0.177) to 0.124 (0.102–0.145), and also achieved higher predictive performance (AUROC: 0.730 [0.712–0.747]; AUPRC: 0.766 [0.749-0.782]) than the baseline model.

We compared LesionAttn against three established bias mitigation methods: LAFTR[20], ADNet[13], and ShorT[9], using the same model architecture of RANN. See more implementation details in the Method section under "Experiment settings". Across internal (HAM testing set) and external (BCN dataset) cohorts, LesionAttn consistently achieved the best fairness and high predictive performance among all bias mitigation methods. As shown in Figure 2B, in the HAM dataset, LesionAttn achieved the lowest EO of 0.062 (0.042-0.082), outperforming other bias mitigation methods (ADNet: 0.099 [0.076-0.123]; LAFTR: 0.090 [0.052-0.128]; ShorT: 0.077 [0.043-0.111]), with comparable predictive performance. In the BCN dataset, all bias mitigation methods achieved better predictive performance relative to the baseline, but only LesionAttn improved the fairness (LesionAttn: 0.124 [0.102–0.145]; ADNet: 0.180 [0.171-0.188]; LAFTR: 0.178 [0.163-0.194]; ShorT: 0.154 [0.140-0.167]).

**Impacts of attention-based training**

To assess the effect of attention-guided training, we examined attention maps from LesionAttn, other bias mitigation methods, and the baseline model. Spatial alignment of models' attention and the reference lesion regions was quantified using intersection over union (IoU). As shown in Figure 3 (left panel), LesionAttn achieved the highest median IoU (0.802) and the lowest variability (standard deviation [SD]: 0.199) over the testing set, outperforming other bias mitigation methods and the baseline (ShorT: median 0.214, SD 0.249; LAFTR: median 0.243, SD 0.270; ADNet: median 0.578, SD 0.288; baseline: median 0.421, SD 0.278).

As shown in Figure 3 (right panel), attention patterns varied across bias mitigation methods. LAFTR typically highlighted non-lesion regions in its attention maps (i.e., bright areas for the lesion regions and dark areas in the non-lesion regions). Similar attention patterns also occasionally appeared in ADNet, ShorT, and the baseline, particularly for the male negative cases. In contrast, LesionAttn consistently concentrated attention on lesion areas, regardless of gender or malignancy status, suggesting a more clinically aligned model reasoning.

In addition, restricting the image to lesion regions alone was not as effective as the attention guidance employed by LesionAttn. We compared LesionAttn with a variant that replaced the learnable attention map with static lesion masks, termed LesionOnly. As shown in Figure 4, LesionOnly achieved low EO values in both datasets at the cost of markedly lower prediction performance. In HAM, LesionOnly yielded an AUROC of 0.624 (0.591-0.657) and an AUPRC of 0.271 (0.253-0.290), far below both the baseline and LesionAttn. Despite LesionOnly's reduction to EO (0.073 [0.064-0.081]) compared to the baseline model, it was still outperformed by LesionAttn. In BCN, LesionOnly significantly reduced gender disparity (0.063 [0.052-0.074]), but collapsed to an AUROC of 0.537 (0.450-0.490) and an AUPRC of 0.586 (0.547-0.626). These results indicate that hard cropping reduced gender bias but destroyed the spatial context essential for accurate prediction.

## Discussion

While AI shows promise in early skin cancer detection, its fairness across gender subgroups remains insufficiently studied. In this study, we introduced LesionAttn, an attention-guided algorithm that integrates clinical knowledge into model training and incorporates fairness in model selection. Using two large-scale dermatological datasets, LesionAttn effectively mitigated algorithmic gender bias while maintaining strong predictive performance, compared to fairness-unaware model development. This study also highlights that incorporating human expertise into AI design can enhance both model performance and fairness, underscoring the value of domain knowledge as a key driver of equitable and trustworthy AI systems.

We found that the baseline model consistently exhibited lower TPRs in female patients across both internal and external cohorts. Because delayed diagnosis of melanoma can lead to rapid disease progression and worse prognosis, prioritizing high TPR is critical for triage and AI-assisted early detection.[30,31] These disparities suggest that female patients may face a greater risk of missed or delayed diagnoses, raising concerns about the equitable deployment of AI models in dermatology. As shown in Figure 3, the baseline model tends to focus on background skin regions rather than lesions when not explicitly guided[32-34]. This misdirected attention may not be incidental: background skin can encode gender-related cues, and overreliance on such features likely contributes to the observed TPR gap.

LesionAttn addresses this issue by algorithmically aligning model focus with lesion regions used by clinicians during diagnosis. It incorporates lesion information during training but does not require it at inference, as demonstrated on the HAM test set (i.e., internal evaluation) and the BCN dataset (i.e., external evaluation). This design significantly reduces gender bias without compromising predictive accuracy, outperforming other bias mitigation algorithms in both internal and external evaluations. As shown in Figure 2, other bias mitigation methods often lacked generalizability, with fairness gains failing to translate across datasets, highlighting the robustness of LesionAttn in real-world settings.

Among the comparative bias mitigation methods, ShorT also improved fairness while preserving performance by discouraging the model from exploiting visual shortcuts.[9] As shown in Figure 3, its attention maps were more diffusely distributed, likely reducing bias by minimizing dependence on spurious background cues of demographics.[32,35] However, ShorT offers no positive guidance on where the model should focus for its diagnosis, as the model learns only what to avoid rather than what to prioritize, thereby limiting clinical interpretability. In contrast, LesionAttn adopted a fundamentally different and clinically interpretable strategy by explicitly directing model attention to lesion regions known to be diagnostically relevant. This approach not only proved more effective but also provided clear insight into where and how the model achieved its fairness status.

This study highlights the potential of embedding clinical knowledge (e.g., the localization of disease-relevant features) into AI model design to advance both fairness and accuracy. Conventional bias mitigation methods often impose an inherent trade-off between these two[23,24], as no extra information is inserted to guide model learning. In contrast, LesionAttn integrates simple yet clinically meaningful priors directly into model training, aligning attention with diagnostically relevant lesion regions and providing a shared direction for optimizing both performance and fairness. Rather than eliminating the fairness–performance trade-off, it shifts the frontier forward, demonstrating that domain-informed AI can achieve a more favorable balance between these competing objectives.[36]

Our comparative analysis also shows that solely cropping images to remove background skin, as done in the LesionOnly variant, yields better fairness but at the cost of catastrophic loss in prediction performance. This underlines the importance of maintaining contextual features rather than enforcing rigid attention constraints.[37] LesionAttn strikes this balance through soft-guidance attention guidance, preserving spatial relationships while learning to prioritize medically relevant areas.[38]

There are limitations to this study. First, while we acknowledge non-binary gender identities, the datasets used were annotated only with binary gender labels, limiting the granularity of subgroup analysis. Second, LesionAttn requires lesion annotations during training, which may introduce annotation burden; however, once trained, the model can operate independently of these labels during inference.

This study reveals gender bias in AI-based skin cancer diagnosis and demonstrates that aligning model attention with clinically relevant lesion regions can mitigate these disparities without compromising accuracy. LesionAttn exemplifies how incorporating clinical expertise into algorithm design promotes fairness, interpretability, and clinical reliability. This approach offers a generalizable approach for building equitable medical AI systems and fosters interdisciplinary collaboration toward the responsible integration of AI in healthcare.

# Methods

**Dataset**

In this study, we focused on predicting the malignancy of skin lesions from dermoscopic images.[5,39] The HAM dataset comprises 10,015 dermoscopic images collected over a two-decade period from the Medical University of Vienna (Austria) and a dermatology clinic in Queensland (Australia).[25] The BCN dataset comprises 19,424 dermoscopic images acquired between 2010 and 2016 at the Hospital Clinic in Barcelona, of which 12,411 images with a confirmed outcome of interest were included in this study.[26] Both datasets cover various skin conditions. In HAM, basal cell carcinoma (BCC), melanoma (MEL), and actinic keratoses/intraepithelial carcinoma (AKIEC) were labeled as malignant. In BCN, BCC, MEL, actinic keratoses (AK), and squamous cell carcinoma (SCC) were considered malignant.

Demographic and clinical characteristics stratified by gender are summarized in Table 1. In both datasets, male patients were older on average and had higher malignancy rates than female patients. To assess model robustness and generalizability, we conducted both internal and external validations. HAM was randomly split into 60% training, 20% validation, and 20% internal testing sets. The entire BCN dataset was reserved for external testing to evaluate performance in an independent clinical setting.

**Residual Attention Neural Network (RANN) for skin cancer diagnosis**

The RANN model is a widely adopted model architecture in skin cancer diagnosis[27-29], which integrates attention maps into a residual learning framework (Figure 1A). We trained the RANN model following established protocols for AI-based skin cancer classification.[27-29] As shown in Figure 1A, the model processes an input dermoscopic image $X$, where the attention is computed via:
$$A = \text{Softmax}(\text{Conv}(X)),$$
with $\text{Conv}(X)$ denoting a convolution transformation applied to $X$ and the activation function Softmax ensuring a spatial probability distribution. The attention-weighted image is then computed as:
$$X_A = A \odot X,$$
and passed through residual blocks for classification. This architecture allows the network's attention to be visualized, facilitating interpretability and clinical alignment.[28]

**LesionAttn algorithm for bias mitigation**

We proposed LesionAttn to improve gender fairness in skin cancer diagnosis while preserving predictive performance. This bias mitigation approach employs two key mechanisms (Figure 1B): (1) soft-guided attention alignment with lesion annotations, and (2) Pareto-based multi-objective model selection.

*Soft-guidance of model attention*

To guide the model's attention to clinically relevant lesion regions, we maximize the alignment between attention maps and lesion masks using cosine similarity. This encourages the model to prioritize features within lesion regions and reduce reliance on possibly gender-correlated background features. Given a lesion mask $M$ and the attention map $A$, the alignment is quantified as:

$$\text{Cos}(M, A) = \frac{M \cdot A}{\|M\| \, \|A\|},$$

where $\cdot$ denotes the dot product, and $\|\cdot\|$ denotes the $L_2$ norm. While focusing on lesions is beneficial, maintaining the contextual features is also crucial to fully leverage the model's capability[38]. To preserve contextual skin features, we introduce a softened mask $M_s$:

$$M_s = \rho + (1 - \rho) \cdot M,$$

where $\rho \in [0, 1]$ controls attention to the background. The attention loss is then defined as:

$$l_A = 1 - Cos(M_s, A)$$

as a component of the overall loss function for model training.

*Pareto frontier for model selection*

Model selection in fairness-aware modeling often involves balancing between competing objectives, such as maximizing predictive performance while minimizing algorithmic bias.[23,24] To address this, LesionAttn employed the Pareto Frontier (PF) algorithm[40], a widely used approach in multi-objective optimization, to identify models representing optimal trade-offs between the objectives. Unlike approaches that combine fairness and accuracy into a single composite loss—forcing one objective to be optimized at the expense of the other[23,24]—the PF framework preserves the independence of both objectives.[40] This allows the model to explore solutions that simultaneously achieve high predictive performance and fairness without imposing artificial weighting.

In our formulation, we simultaneously maximized predictive performance ($P_{\text{pred}}$, measured by AUROC) and fairness ($P_{\text{fair}}$, quantified by $1 - EO$). A model $f$ lies on the Pareto frontier (i.e., Pareto-optimal) if there exists no other model $f'$ that performs at least as well in both objectives and strictly better in one:

$$P_{\text{pred}}(f') \geq P_{\text{pred}}(M) \text{ and } P_{\text{fair}}(M') \geq P_{\text{fair}}(M),$$

with at least one strict inequality.

**Attention analysis**

To examine the attention behavior, we extracted and visualized attention maps from RANN models trained with the baseline, LAFTR, ADNet, ShorT, and LesionAttn. Attention maps were obtained from the final attention layer and overlaid on the input images to compare the models' regions of focus. Quantitatively, the attention-lesion alignment was measured by the IoU values between each model-generated attention map and the corresponding lesion mask annotated by dermatology experts. For illustration, we randomly selected four cases from the HAM testing set for examples: male negative, female negative, male positive, and female positive, respectively.

**Experiment settings**

To evaluate the effectiveness of LesionAttn, we compared it with three established bias mitigation methods (LAFTR, ADNet, and ShorT) under the same experimental settings. All models shared the same RANN architecture to ensure consistency in comparison. Additionally, to isolate the effect of learnable attention, we constructed a LesionOnly variant of RANN, in which input images were cropped using binary lesion masks rather than learned attention maps. This setup enabled assessment of whether static lesion focus alone could reproduce the fairness improvement achieved by LesionAttn.

All models underwent hyperparameter optimization via grid search, using AUROC on the validation set as the selection criterion. Hyperparameter ranges and final settings are detailed in the Supplementary eTable 1. Each experiment was repeated five times with different random seeds, and results are reported as mean values with 95% confidence intervals.

## Ethics declarations
Ethics approval and consent to participate: Not applicable.

## Author Contributions
Conceptualization: M.Z., M.L., N.L. Method design, experiment, data analysis: M.Z., M.L. Drafting of the manuscript: M.Z., M.L.. Critical revision of the manuscript: M.Z., M.L., N.L. Interpretation of the content: M.Z., M.L., H.Y., Y.N., N.L. Revisions of the manuscript: M.Z., M.L., H.Y., Y.N., N.L. Final read and approval of the completed version: all authors. Overseeing the project: N.L.

## Acknowledgement

This work was supported by the Duke-NUS Signature Research Programme funded by the Ministry of Health, Singapore. Any opinions, findings and conclusions or recommendations expressed in this material are those of the author(s) and do not reflect the views of the Ministry of Health. Tingting Zhu was supported by the Royal Academy of Engineering under the Research Fellowship scheme.

## Competing Interests

The authors declare that there are no competing interests.

**Table 1.** Demographics of each gender group of HAM and BCN.

|  | Average Age | Positive Rate | Count |
|---|---|---|---|
| HAM (Internal) |  |  |  |
| Male | 54.545 | 0.227 | 5,406 |
| Female | 48.712 | 0.160 | 4,552 |
| Total | 51.864 | 0.196 | 10,015 |
| BCN (External) |  |  |  |
| Male | 59.121 | 0.586 | 6,499 |
| Female | 54.137 | 0.428 | 5,840 |
| Total | 56.762 | 0.511 | 12,411 |

**Figure 1.** Illustration of the AI-based skin cancer diagnosis and the LesionAttn algorithm.

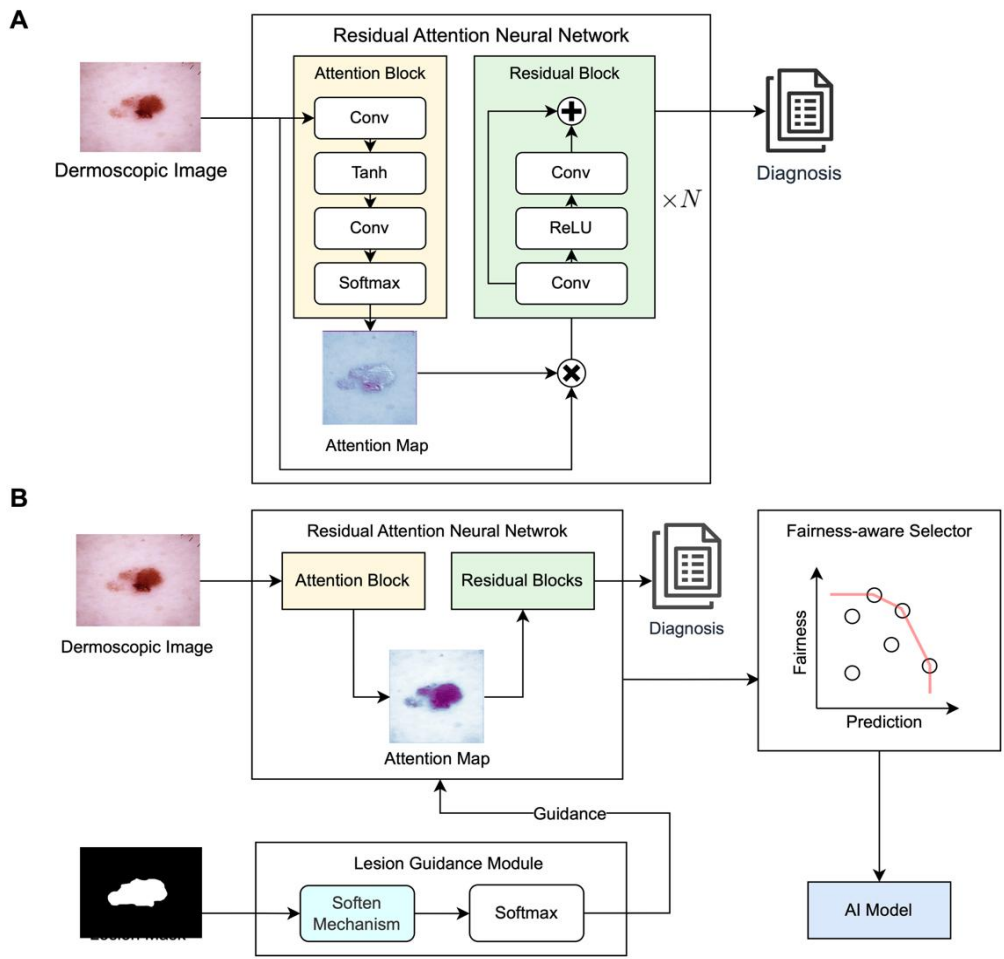

**A.** The detailed architecture of the Residual Attention Neural Network (RANN), including an attention block for generating attention maps and multiple residual blocks for inference. **B.** The workflow of the LesionAttn. The Soften Attention Guidance Module guides the attention map during training. This module encourages the AI model to focus on lesions by increasing the cosine similarity between the attention map and the softened lesion mask through the loss function. The Fairness-aware Selector is responsible for selecting the AI with optimal fairness and prediction performance based on the Pareto Frontier algorithm.

**Figure 2.** Disparity in the baseline model and comparison of five bias mitigation methods

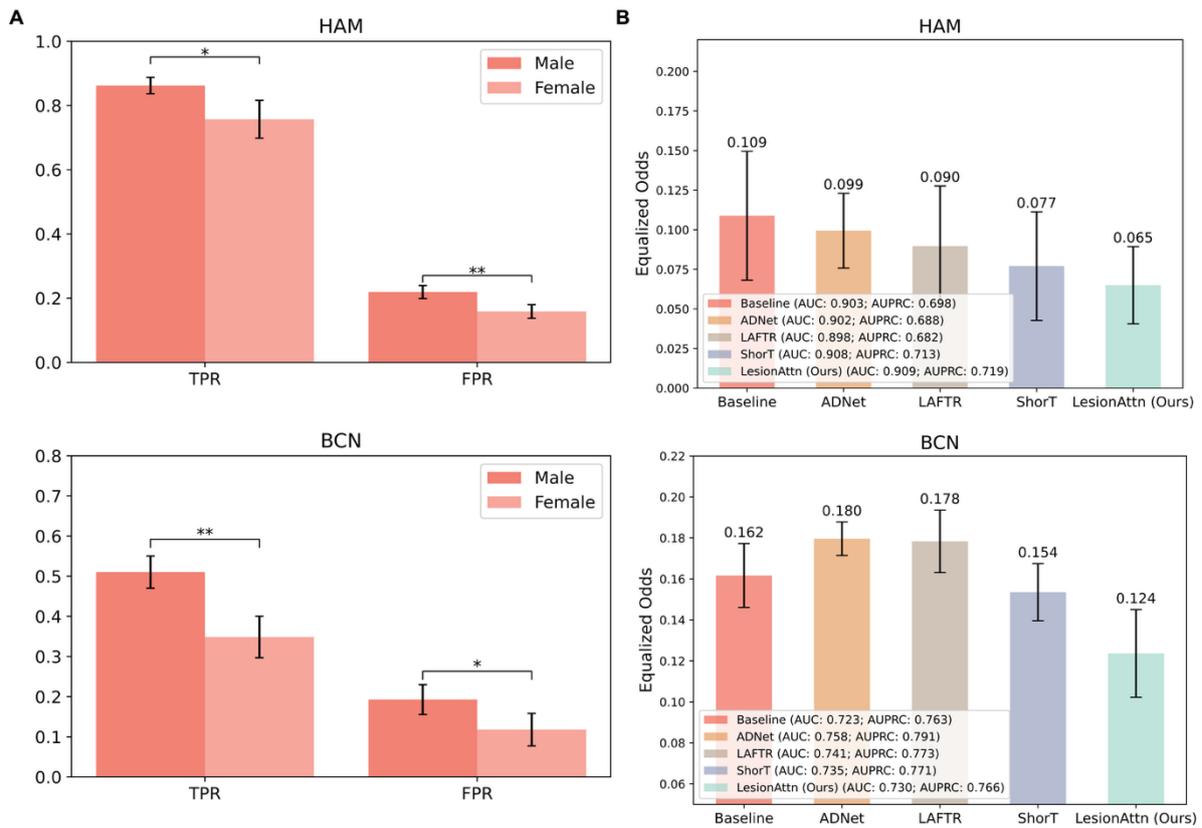

**A.** Gender disparity of the baseline model on HAM and BCN datasets. The fairness is evaluated by comparing the difference in TPR and FPR between male and female groups. **B.** Comparison of bias mitigation methods.

**Figure 3.** Visualization of input images and their corresponding attention maps generated by RANN models trained with Baseline, LAFTR, ADNet, ShorT, and LesionAttn.

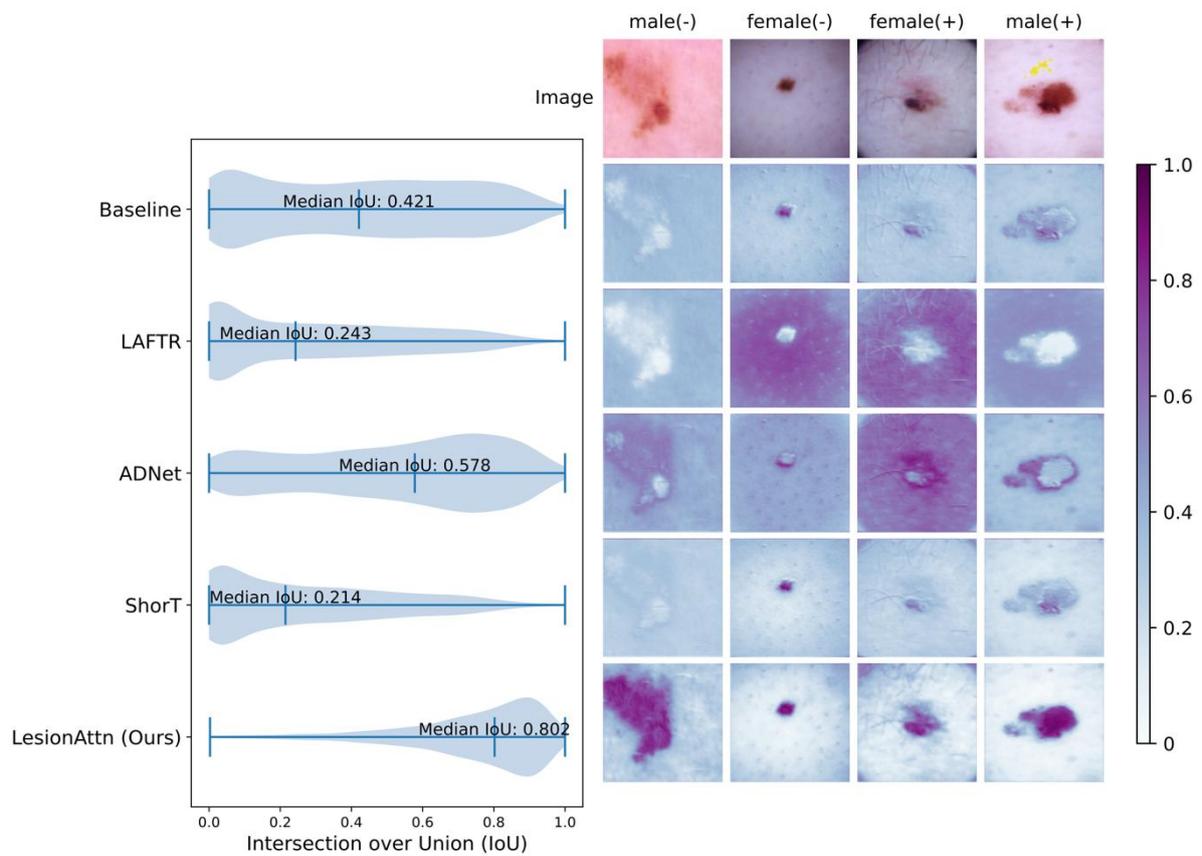

Higher heatmap intensity indicates greater attention to one pixel, while lower intensity indicates less attention to one pixel.

**Figure 4.** Comparison between LesionAttn and LesionOnly.

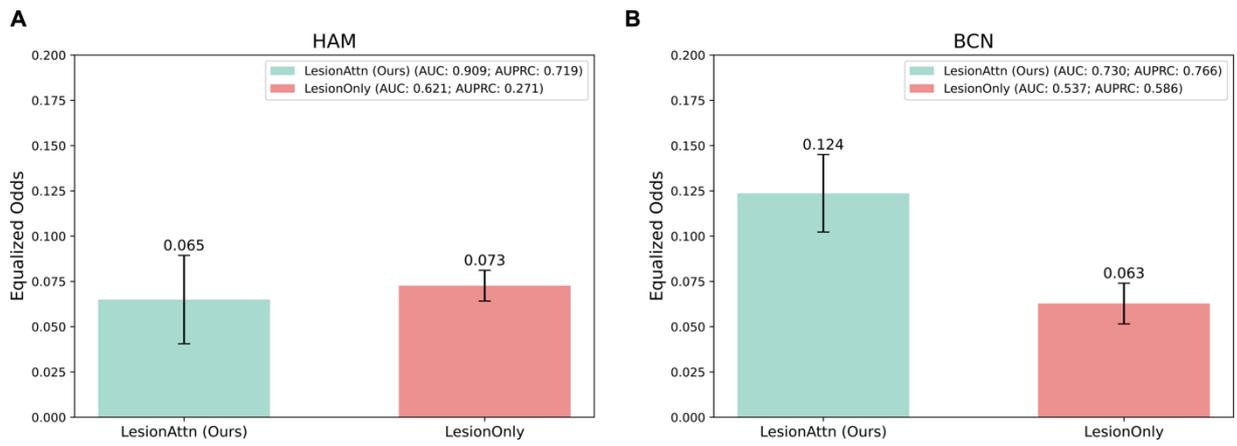

# Supplementary Materials

**eTable 1.** Hyper-parameters for each algorithm

| Algorithm | Parameters | Search Space | Optimal Value |
|---|---|---|---|
| The baseline | | | |
| | Learning Rate (LR) | $[1 \times 10^{-5}, 1 \times 10^{-4}, 1 \times 10^{-3}]$ | $1 \times 10^{-5}$ |
| LAFTR[20] | | | |
| | LR | $[1 \times 10^{-5}, 1 \times 10^{-4}, 1 \times 10^{-3}]$ | $1 \times 10^{-5}$ |
| | Discriminator LR | $[1 \times 10^{-5}, 1 \times 10^{-4}, 1 \times 10^{-3}]$ | $1 \times 10^{-5}$ |
| | Adversarial Loss Coefficient | [0.5, 1.0, 2.0] | 0.5 |
| | Reconstruction Loss Coefficient | [0.5, 1.0, 2.0] | 0.5 |
| ADNet[13] | | | |
| | LR | $[1 \times 10^{-5}, 1 \times 10^{-4}, 1 \times 10^{-3}]$ | $1 \times 10^{-5}$ |
| | Discriminator LR | $[1 \times 10^{-5}, 1 \times 10^{-4}, 1 \times 10^{-3}]$ | $1 \times 10^{-5}$ |
| | Adversarial Loss Coefficient | [0.5, 1.0, 2.0] | 0.5 |
| ShorT[9] | | | |
| | LR | $[1 \times 10^{-5}, 1 \times 10^{-4}, 1 \times 10^{-3}]$ | $1 \times 10^{-5}$ |
| LesionAttn | | | |
| | LR | $[1 \times 10^{-5}, 1 \times 10^{-4}, 1 \times 10^{-3}]$ | $1 \times 10^{-5}$ |
| | Attention Loss Coefficient | $[1 \times 10^{-5}, 1 \times 10^{-4}, 1 \times 10^{-3}]$ | 0.5 |
| | Soften Value | [0.0, 0.1, 0.2, …, 0.9] | 0.7 |

The model is trained using the Adam optimizer, with $\beta_1 = 0.9, \beta_2 = 0.999$. The learning rate will decay to 99% of its current value after every 10 training steps. Training was conducted for 100 epochs per algorithm, with early stopping implemented if no improvement in performance was observed within 10 epochs.

**eFigure 1.** ROC curves of models

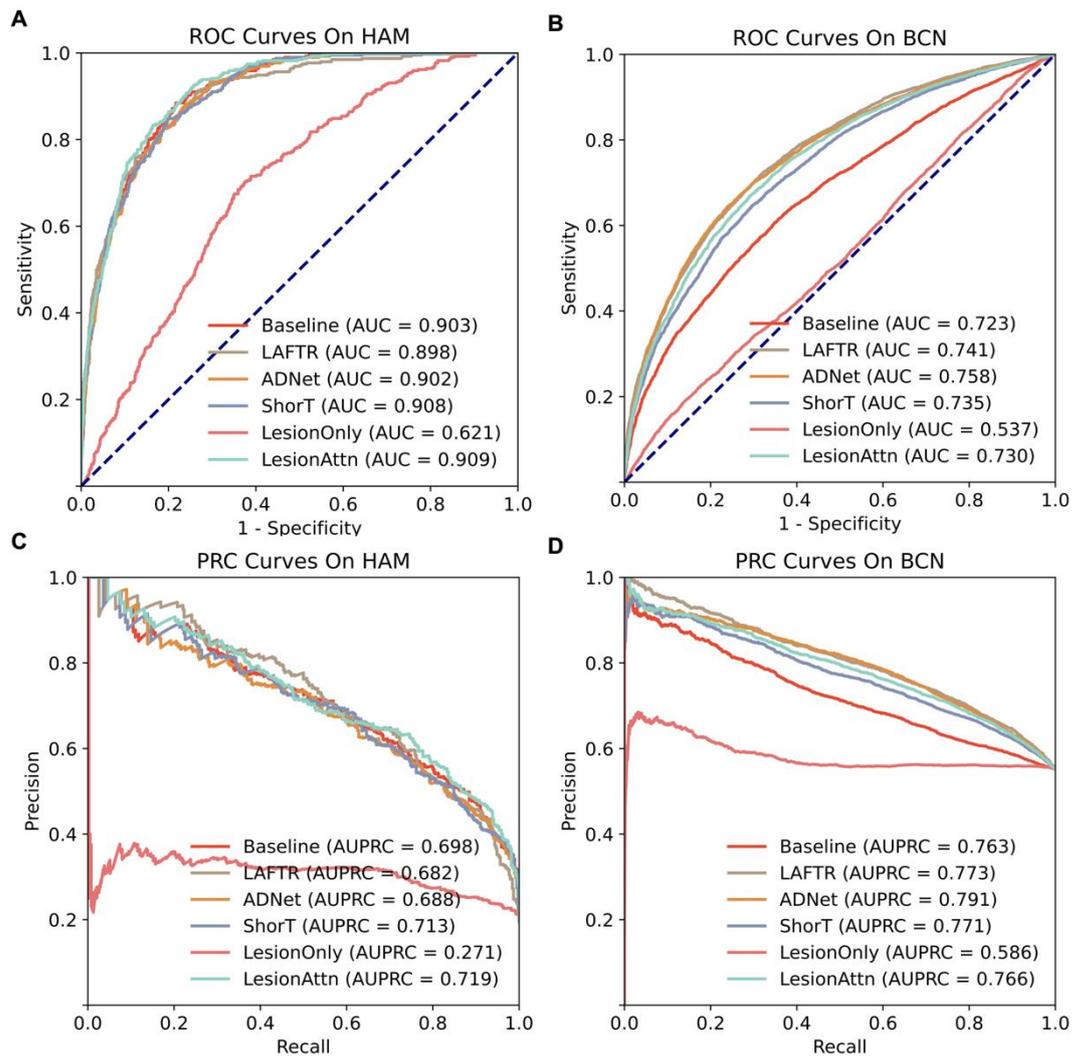